\documentclass[a4paper]{article}
\usepackage{listings}
\usepackage{xcolor}
\usepackage{multirow}
\usepackage{jheppub}
\usepackage{dcolumn}
\usepackage[english]{babel}
\usepackage[utf8x]{inputenc}
\usepackage[T1]{fontenc}
\usepackage{nicefrac}

\usepackage{multirow}

\usepackage{amsmath}
\usepackage{afterpage}
\usepackage{graphicx, subcaption}
\usepackage[colorinlistoftodos]{todonotes}
\usepackage[colorlinks=true]{hyperref}

\usepackage{wasysym}

\newcommand{\be}{\begin{equation}}
\newcommand{\ee}{\end{equation}}
\newcommand{\bea}{\begin{eqnarray}}
\newcommand{\eea}{\end{eqnarray}}

\newcommand{\Earth}{{\scalebox{0.4}{$\bigoplus$}}}

\begin{document}

\vspace*{30mm}

\begin{center}
{\LARGE \bf Bounding  Quantum Dark Forces
}
\par\vspace*{20mm}\par

{\large \bf Philippe Brax$^a$, Sylvain Fichet$^b$, Guillaume Pignol$^c$}

\bigskip

{\em $^a$ Institut de Physique Th\'{e}orique, Universit\'e Paris-Saclay, CEA, CNRS, F-91191 Gif/Yvette Cedex, France}
\\
{\em $^b$ ICTP-SAIFR \& IFT-UNESP, R. Dr. Bento Teobaldo Ferraz 271, S\~ao Paulo, Brazil}
\\
{\em $^c$ Laboratoire de Physique Subatomique et de Cosmologie, Universit\'e Grenoble-Alpes, CNRS/IN2P3, Grenoble, France  }
\vspace*{5mm}

\vspace*{15mm}

{  \bf  Abstract }

\end{center}
\vspace*{1mm}

\noindent
\begin{abstract}


Dark sectors lying beyond the Standard Model  and  containing  sub-GeV particles which are bilinearly coupled to nucleons would induce quantum forces of the Casimir-Polder type in ordinary matter.
 Such new forces can be tested by a variety of experiments over many orders of magnitude.  We provide a generic interpretation of these experimental searches  and apply it to a sample of forces from dark scalars behaving as $1/r^3$, $1/r^5$, $1/r^7$  at short range.
The landscape of constraints on such quantum forces  differs from the one of modified gravity with Yukawa interactions, and features in particular strong  short-distance bounds from molecular spectroscopy and neutron scattering.

\end{abstract}

\vspace{5.2cm }
\noindent
{\em E-mail:\\
philippe.brax@ipht.fr \\ sylvain@ift.unesp.br\\ guillaume.pignol@lpsc.in2p3.fr
\\
}

\noindent
\newpage

\section{Introduction}\label{sec:intro}

When  going beyond  the Standard Model (SM) of particle physics, it is natural to imagine the existence of other light particles, which would have been so far elusive because of their weak or vanishing interactions with the SM particles.
 Such  speculations on   \textit{dark sectors} could be simply driven by theoretical curiosity although  there are   more concrete motivations coming from  two striking observational facts:
Dark Matter and  Dark Energy. In both cases,  theoretical constructions elaborated  to explain one or both of these fundamental aspects of the Universe tend to assume the existence of dark sectors of various complexity.

Among the many possibilities for the content of the dark sector, our interest in this work lies in dark particles with masses below the GeV scale, where  Quantum ChromoDynamics  (QCD) reduces to an effective theory  of nucleons. Would a light scalar couple to nucleons, it would induce a fifth force  of the form $V = \alpha e^{-r/\lambda}/r$, with $\lambda=\hbar/ mc$ being the Compton wavelength of the scalar and $m$ its mass. The presence of such Yukawa-like force is sometimes dubbed  ``modified gravity''.  Experimental searches for such fifth forces between nucleons extend from nuclear to astronomic scales and lead  to a landscape of exclusion regions, see summary plots in \cite{Fischbach:1999bc,Adelberger:2003zx,Adelberger:2009zz,Antoniadis:2011zza,Salumbides:2013dua}.

As noted in \cite{Fichet:2017bng}, even in the absence of a light boson linearly coupled to nucleons, other fifth forces can still arise from the dark sector whenever a sub-GeV particle of any spin is  bilinearly coupled to nucleons.   Such forces would arise from the double exchange of a particle and are thus fundamentally quantum. Moreover, in order to take into account retardation effects, such forces have to be computed within relativistic quantum field theory. This kind of computation has been first done by Casimir and Polder for polarizable particles~\cite{CasPol}, and by Feinberg and Sucher for neutrinos~\cite{Feinberg:1968zz}. We will refer to such quantum forces as Casimir-Polder forces.

There is a variety of motivations for having a particle of the dark sector coupling bilinearly to nucleons. The dark particle can be for instance  charged under a symmetry of the dark sector, can be a symmetron from a dark energy model, or simply a dark fermion sharing a contact interaction with nucleons.  Such $Z_2$ symmetry can be also needed in order to explain the stability of Dark Matter.


In the presence of forces which do not have a Yukawa-like behaviour, as is the case of the Casimir-Polder forces we focus on, the  landscape of fifth force searches is expected to  change drastically. A thorough investigation of the experimental fifth force searches  becomes then mandatory in order to put bounds on such extra forces in a consistent manner, and thus on the underlying dark particles.

 This requires revisiting  each of the experimental results, a task that will be performed in this paper.
In Sec.~\ref{sec:EFT}, we consider Casimir-Polder forces focussing on the case  of a scalar with various effective interactions with nucleons. General features of  Casimir-Polder forces are then derived in Sec.\ref{sec:properties}.
A generic interpretation of   the most recent and stringent fifth force searches, valid for arbitrary potentials, is given in Sec.\ref{sec:forces}. The exclusion regions will be displayed and discussed in Sec.~\ref{sec:results}.

We emphasize that our approach to constrain dark sectors relies only on virtual dark particles, and is thus independent on whether or not the dark particle is stable. The case where the dark particle is stable and identified as Dark Matter has been treated in a dedicated companion paper, Ref.~\cite{Fichet:2017bng}.
 Searches for dark sectors via loops of virtual dark particles include Refs.~\cite{Fichet:2016clq,Fichet:2017bng,Voigt:2017vfz}, and are yet under-represented in the literature.

\section{Casimir-Polder forces from a dark scalar}\label{sec:EFT}


There are many reasons for which the dark sector could feature a scalar with a  $Z_2$ symmetry with respect to the Standard Model sector. If such a scalar is charged under a new symmetry such as a $U(1)_X$ charge while the SM fields are not, the scalar should interact with the SM via bilinear operators. The scalar  can also be the pseudo-Nambu-Goldstone boson (pNGB) of an approximate global symmetry, in which case it couples mostly with derivative couplings to the nucleons.
Theories of modified gravity can also feature light scalars  with a bilinear coupling to the stress-energy tensor \cite{Copeland:2006wr}. While the properties of these scalars are often considered to be modified by some screening mechanism, it is certainly  relevant to consider scenarios where screening is negligible or absent. This is the most minimal possibility, and can also serve as a reference for comparison with the screened models. Moreover for models like the symmetrons, screening does not happen in vacuum.

It is convenient to use an effective field theory (EFT) approach to describe the interactions of the dark particle. All the measurements we consider occur well below the quantum chromodynamics (QCD) confinement scale, hence we can readily write down effective interactions with nucleons. The operators we consider have the form ${\cal O}_{\rm nuc}{\cal O}_{\rm DS}$, where ${\cal O}_{\rm nuc}$ is bilinear in the nucleon fields and ${\cal O}_{\rm DS}$ is bilinear in the dark sector field. ${\cal O}_{\rm nuc}$ has in principle a $\bar N \Gamma^A  N$ structure, where $\Gamma^A$ can have any kind of Lorentz structure.
 In the limit of unpolarized non-relativistic nucleons, only the interactions involving ${\cal O}_{\rm nuc}=\bar N N,\bar N \gamma^0  N$ are relevant, the other being either canceled  by averaging over  nucleon spins or suppressed by powers of $m_N^{-1}$.

 In this paper we focus on the exchange of a dark scalar. The exchange of dark fermions and dark vectors, either self-conjugate or complex, have been treated in  \cite{Fichet:2017bng}, and details of the calculations  for all these cases are given in App.~A. Here we focus on  three types of effective interactions, ${\cal L}={\cal L}_{\rm SM} + {\cal O}_i$, with
\bea
{\cal O}^0_a=\frac{1}{\,\Lambda} \bar N N \frac{\phi^2}{2} \,, \quad
{\cal O}^0_b=  \frac{1}{\,\Lambda^2}\, \bar N \gamma^\mu N \phi^* i\overleftrightarrow\partial_\mu  \phi \,,
\quad
{\cal O}^0_c=\frac{1}{\,\Lambda^3} \bar N N \frac{(\partial^\mu\phi)^2}{2} \,. \label{eq:Leff}
\eea
We  assume that only one of these operators is turned on at a time. In the ${\cal O}^0_{a,c}$ cases, we assume a real scalar, while for  ${\cal O}^0_b$ we assume a complex scalar.
The ${\cal O}^0_a$ interaction corresponds to the case of a symmetron, the ${\cal O}^0_b$ interaction is typically the one generated from a heavy $Z'$ exchange, and the ${\cal O}^0_c$ would occur if the scalar is the pNGB of a hidden global symmetry. In the last case,  as the pNGB mass explicitly breaks the shift symmetry, an interaction of the form $\frac{m^2}{\Lambda^2} {\cal O}^0_a$ could also be present, however its effect would be negligible at short distance hence we do not take it  into account. 
 Similar calculations have been performed for disformal couplings in \cite{Kaloper:2003yf,Brax:2014vva}.

 Higher dimensional operators are in principle present in the effective Lagrangian, and are suppressed by higher powers of either $\Lambda$ or $\Lambda_{\rm QCD}$. The EFT is valid for momenta below  $\min{(\Lambda, \Lambda_{\rm QCD})}$ when coupling constants are $O(1)$ in the UV theory.
We will assume a universal coupling to protons and neutrons---all our results are easily generalized for non-universal couplings. Also, for simplicity, we do not  consider the dark particle coupling to electrons. Including the coupling to electrons would lead typically to stronger forces and thus to enhanced limits.

\begin{figure}[t]
  \centering
  \includegraphics[scale = 0.6,trim={0 0 0 0},clip]{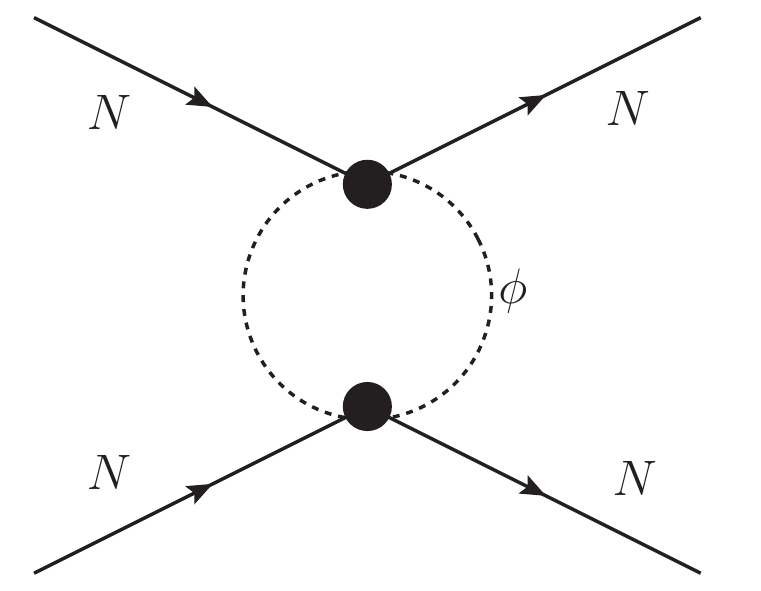}
   \caption{
The exchange of two scalars inducing a force between the nucleons.  }
  \label{fig:loop}
\end{figure}

As a result of the   ${\cal O}_{a,b,c}$ interactions, nucleons can exchange \textit{two} scalars as shown in the Feynman diagram of Fig.~\ref{fig:loop}. This Feynman diagram induces   a Casimir-Polder force (\textit{i.e.} a relativistic van der Waals force) between the nucleons. The forces induced by the ${\cal O}_{a,b,c}$ operators   have been computed in \cite{Fichet:2017bng} and are given by the potentials
\bea \nonumber
V_a &=&  -\frac{1}{ 32 \pi^3\, \Lambda^2 } \frac{m}{r^2} K_1(2m r)\,, \quad \quad V_b = \frac{1}{8 \pi^3\Lambda^4  } \frac{m^2}{r^3} K_2(2m r)\,, \\
 V_c &=& -\frac{1}{32 \pi^3\,\Lambda^6\, r} \left(\left(\frac{30m^2}{r^4}+\frac{6m^4}{r^2}\right) K_2(2m r)+ \left(\frac{15m^3}{r^3}+\frac{m^5}{r}\right) K_1(2m r) \right)
     \,, \label{eq:forces}
\eea
where $K_i$ is the $i$-th modified Bessel function of the second kind. The $V_a$ force is consistent with a previous calculation of \cite{Grifols:1996fk} after matching to our conventions.

The main steps of the general calculation are as follows. One first calculates the amplitude corresponding to the diagram in Fig.\,\ref{fig:loop}. In order to calculate loop amplitudes in the EFT, dimensional regularization has to be used in order not to spoil the EFT expansion. The one-loop amplitudes can be decomposed over the basis
\be
f_n=\int_0^1 dx (x(1-x))^n \log\left(\frac{\Delta}{\Lambda^2}\right)\label{eq:fn}
\ee
where $\Delta=m^2-x(1-x)q^2$.  $\Lambda$ is the scale at which the effective theory is matched on to the UV theory, and is also the scale at which the EFT breaks down.

 Then one takes the non-relativistic limit of the amplitude and identify the scattering potential $\tilde V$ as  \be
i{\cal M} = -i \tilde V({\bf |q|})   4m_N^2 \delta^{s_1s_1'}\delta^{s_2s_2'}\,, \label{eq:NRV}
\ee
where $s_{1,2}$ $(s'_{1,2})$ corresponds to the  spin polarization of each ingoing (outgoing) nucleons. The spatial potential is given by the 3d Fourier transform of $\tilde V({\bf |q|})$,
\be
V(r)=\int\frac{ d^3 { \bf q} }{(2\pi)^3} \tilde V(|{\bf q}|) e^{i { \bf q}\cdot {\bf r}}= \frac{-i}{(2\pi)^2\,r} \int_{-\infty}^{\infty} d\rho \rho \tilde V(\rho) e^{i \rho r} \,,
\ee
where $r=|{\bf r}|$ and the momentum has been extended to the complex plane in the last equality, $\rho\equiv |{\bf q}| $. Using standard  complex integration one obtains
\be
 V(r)=\frac{-i}{(2\pi)^2\,r} \int_{i2m}^{i\infty} d\rho \rho [\tilde V] e^{i \rho r}  = \frac{i}{(2\pi)^2\,r} \int_{2m}^{\infty} d\lambda \lambda [ \tilde  V] e^{- \lambda r} \label{eq:V2}
\ee
where $[V]$ is the discontinuity from right to left across  the positive imaginary axis, $[V]=V_{\rm right}-V_{\rm left}$, and one has defined $\rho=i\lambda$. Notice that $\lambda$ can also be understood as  $\sqrt{t}$, the square root of the $t$ Mandelstam variable extended to the complex plane.
The discontinuities $[f_n]$ needed    to compute the Casimir-Polder force via  Eq.~\eqref{eq:V2} are given in  Appendix A.

In the case of the scalar dark particle exchanged via the ${\cal O}_a$, ${\cal O}_b$ or ${\cal O}_c$ operators, the amplitudes are given in App.~B.  The discontinuities needed to calculate the $V_{a,b,c}$ potentials are
\bea
 \left[f_0\right]&=&i\pi\frac{2}{\lambda} \sqrt{\lambda^2-4m^2} \,, \\ \nonumber
 \left[f_1\right] &=&i\pi\frac{2m^2+\lambda^2}{3\lambda^3} \sqrt{\lambda^2-4m^2} \,, \\ \nonumber
 \left[f_2\right] &=&i\pi\frac{6m^4+2m^2\lambda^2+\lambda^4}{15\lambda^5} \sqrt{\lambda^2-4m^2} \,.
 \eea
The discontinuity of the nonrelativistic scattering potentials for  the three diagrams considered above are
\bea
\left[ \tilde V_a\right]&=& \frac{\left[f_0\right]}{32 \pi^2\, \Lambda^2} \,, \\ \nonumber
\left[ \tilde  V_b\right]&=& \frac{m^2 \left[f_0\right]-\lambda^2 \left[f_1\right]}{8 \pi^2\, \Lambda^4} \,, \\ \nonumber
\left[\tilde  V_c\right]&=&  \frac{(6m^4 +m^2\lambda^2) \left[f_0\right] +
(24m^2 \lambda^2 +\lambda^4) \left[f_1\right]+
20\lambda^4\, \left[f_2\right]
}{64   \pi^2\, \Lambda^6} \,.\\ \nonumber
\eea

%
At short distance $mr \ll 1$ the forces behave as
\be
V_a = -\frac{1}{64\pi^3\,\Lambda^2\,r^3}\,,\quad
V_b = \frac{1}{16\pi^3 \Lambda^4\,r^5}\,,\quad
V_c = -\frac{15}{32\pi^3 \Lambda^6\,r^7}\,,
\ee
while at long distance $mr \gg 1$ the forces go as
\be
V_a = -\frac{\sqrt{m}\, e^{-2m r} } {64 \pi^{5/2}\Lambda^2\, r^{5/2} } \,,\quad
V_b = \frac{m^{3/2} e^{-2m r} } {16 \pi^{5/2} \Lambda^4 \, r^{7/2} }\,,\quad
V_c = -\frac{m^{9/2} e^{-2m r} } {64 \pi^{5/2}\Lambda^6 \, r^{5/2} }\,.
\ee
As sketched in \cite{Fichet:2017bng}, the broad features of these forces can be understood from general principles. The arguments are given in detail in the next section.

\section{General features of Casimir-Polder forces}\label{sec:properties}

Let us first comment on  the effective theory giving rise to the Casimir-Polder forces. The four-nucleon loop diagrams we consider come from higher-dimensional operators and are thus more divergent than the four-nucleon diagrams from the UV theory lying above $\Lambda$.
This implies that four-nucleon \textit{local} operators (\textit{i.e.}~counter-terms) of the form $(\bar N N)^2$, $(\partial^\mu(\bar N N))^2$, \ldots~are also present in the effective Lagrangian to cancel the divergences which are not present in the UV theory.  The finite contribution from these local operators is fixed by the UV theory at the matching scale,   and is expected to be of same order as the coefficient of the $\log \Lambda$ term in the amplitude by naive dimensional analysis   (this situation is  analog to renormalisation of the non-linear sigma model, see Ref.~\cite{Manohar:1996cq}).
The loop amplitudes have the form \be{\cal M}=F(q^2)+G(q^2) \log\left(\frac{m}{\Lambda}\right)\,,\ee
where $F(q^2)$ is complex, with $F(q^2=0)=0$, and $G(q^2)$ is a real polynomial in $q^2$ (both depend also on $m$, $\Lambda$). The log term is a consequence of the divergence. The log term is real  and contributes to  the running of  local four-nucleon operators. The Casimir-Polder force arises from the branch cut of $F({ q^2})$, and is thus independent of the log term. An experiment measuring only the Casimir-Polder force will have  the advantage of being unsensitive  to these   four-nucleon  operators - which are set by the UV completion and thus introduce theoretical uncertainty. This happens either when the experiment is nonlocal by design (\textit{e.g.}  measuring the force between nucleons at a non-zero distance), or by construction of the observables as we will see in the case of neutron scattering. All the measurements considered in this paper are either fully or approximately unsensitive to local four-nucleon interactions.

The main features of Casimir-Polder forces between two non-relativistic sources  can be understood using dimensional analysis and the optical theorem.
We focus on the double exchange of a particle
having local interactions with the sources, the operators used in Sec.~\ref{sec:EFT} being examples of such scenario. We further assume that the sources are identical---a similar approach applies similary to  different sources. We denote by $X$ the dark particle  exchanged, $\bar X$ its conjugate, $m$ its mass. We use nucleons as source for concreteness. $X$ can take any spin.  The generic operator we consider has the form
\be
{\cal L}\supset \frac{1}{\Lambda^n}{\cal O}(X) \bar N \Gamma^A N \,,
\ee
where $\Gamma^A$ can be any Lorentz structure. When averaging over the nucleon spins, the first non-vanishing Lorentz structures are $\bar N N $ (``scalar channel''), $\bar N \gamma^\mu N $ (``vector channel''), and we will focus on those ones.

Within the above assumptions we obtain  the following properties:
\begin{enumerate}
\item \textit{Sign.}
Operators of the form ${\cal O}(X)\bar N N$  give rise to attractive forces.  Operators of the form ${\cal O}_\mu(X)\bar N \gamma^\mu N$  give rise to repulsive forces.
\item \textit{Short distance.}
 An operator of dimension $n+4$ gives rise to a potential behaving at short distance as
\be
V(r)\propto \frac{1}{r^{1+2n}}\,.
\ee
 \item \textit{Long distance.}
When the square amplitude $|{\cal M}(N\bar N \leftrightarrow X \bar X)|^2$ taken at $\sqrt{s}\sim 2m$
is suppressed by a power $(s-4m^2)^p$ (\textit{i.e.} velocity-suppressed by $v^{2p}$), the long range behaviour of the force is given by
\be
V(r)\propto \frac{e^{-2m r}}{r^{\frac{5}{2}+p}}\,.
\ee
\end{enumerate}

Let us prove the above properties. Property 2 is simply a consequence of dimensional analysis. When $r \ll 1/m $, the potential can be expanded with respect to $r m$ and at first order, $V(r)=V(r)|_{m=0}(1+O( mr))$. In this limit the potential depends only on $r$ and on the effective coupling $1/\Lambda^n$ squared. The potential having dimension 1, it must have a dependence in $1/r^{2n+1}$ so that dimensions match. Notice that this argument applies similarly for the exchange of a single particle (giving then a $1/r$ potential) or for the exchange of an arbitrary number of particles.

For Properties 1 and 3, let us denote the amplitude of interest (Fig.~\ref{fig:loop}) by $i{\cal M}_t$, and introduce  the amplitude $i{\cal M}_s=i{\cal M}(N \bar N  \rightarrow  X^* \bar X^*  \rightarrow N \bar N) $, which is the  $s\leftrightarrow t$ crossing of $i{\cal M}_t$. In order to get some insight  on $i{\cal M}_t$, we can study $i{\cal M}_s$  use crossing symmetry.
The optical theorem applies to $i{\cal M}_s$, with
\bea
&\quad& {\rm Im}({\cal M}_s) = {\rm Im}  \Big( {\cal M} (N \bar N  \rightarrow  X^* \bar X^*  \rightarrow N(q_1) \bar N (q_2)  ) \Big)  \nonumber \\  &=&\frac{1}{2} \sum_{ {\rm polar.}} \int  \frac{d^4 q_1}{(2\pi)^3} \delta(q_1^2)\frac{d^4 q_2}{(2\pi)^3} \delta(q_2^2)(2\pi)^4\delta^{4}(q_1+q_2-q)|{\cal M}(N \bar N \rightarrow X(q_1) \bar X(q_2))|^2 \nonumber \\
&=& \frac{1}{16 \pi} \sqrt{1-\frac{4m^2}{s}}
\sum_{ {\rm polar.}} |{\cal M}(N \bar N \rightarrow X(q_1) \bar X(q_2))|^2 \label{eq:OT}
\eea
where in the last line we use the fact that the amplitude arising from  local interactions (Eq.~\eqref{eq:V2}) depend only on the
 center-of-mass energy $\sqrt{s}$.
The optical theorem is of interest  because ${\rm Im}({\cal M}_t)$ is directly related to the discontinuity of ${\cal M}_t$ over its branch cut, which is precisely the quantity needed to calculate non-relativistic potential. In the formalism of Sec.~\ref{sec:EFT}, we have \be
{\rm Im}({\cal M}_t)= \frac{-1}{2} [\tilde V] 4m_N^2 \delta^{s_1s_2}\delta^{s'_1s'_2}\,. \label{eq:Vtilde}
\ee
It turns out that ${\rm Im}({\cal M}_t)>0$ ($<0$) corresponds to an attractive (repulsive) force.

Let us prove Property 1.
For the scalar channel, the  crossing of $ {\rm Im}({\cal M}_s)$ stays positive,  hence ${\rm Im}({\cal M}_t)>0$ and the force is attractive.  
For the vector channel, we have ${\cal M}(N \bar N \rightarrow X \bar X) \propto J_{\mu,N} J_{X}^\mu$ where the $J_{\mu}$ are vector currents. The square matrix elements takes the form $(J_{\mu,N} J_{\nu,N}) (J_{\mu,X} J_{\nu,X})$. All the $J_{\mu}$ are conserved currents, $J_{\mu} q^\mu=0$.  The $J_{\mu,N}$ can be pulled outside of the integral in  Eq.~\eqref{eq:OT}. Conservation of the $J_{\mu,N}$ currents implies that they project out the  components proportional to  $q_\mu$ of the quantity they are contracted with.
 It follows that
\be
  J_{\mu,N} J_{\nu,N}\sum_{ {\rm polar.}} ( J_{\mu,X} J_{\nu,X})= J_{\mu,N} J_{\nu,N} A(s) ( q^\mu q^\nu - s g^{\mu\nu})\,, \label{eq:As}
\ee
where we have introduced $s=(q_1+q_2)^2$ and $A(s)$ is a positive function. In the non-relativistic limit, one keeps only the $\mu=\nu=0$ components of the nucleon currents, and the projector reduces to $q^\mu q^\nu - s g^{\mu\nu} \sim {\bf q}^2$ ---hence $A(s)$ has to be positive to ensure ${\rm Im}({\cal M}_s)>0$.  The crossing of ${\rm Im}({\cal M}_s)$ gives
\be
{\rm Im}({\cal M}_t)= (\tilde J_{\mu,N} \tilde J_{\nu,N}) A(t)( q^\mu q^\nu - t g^{\mu\nu})\,, \label{eq:At}
\ee
where $\tilde J_{\mu,N}$ denotes the crossed nucleon currents. In the non-relativistic limit we have $\tilde J_{\mu,N} \tilde J_{\nu,N} \sim 4m_N^2 \delta^{\mu 0}\delta^{\nu 0} \delta^{s_1s_2}\delta^{s'_1s'_2}$, $q^0\sim 0 $, $t\sim -{\bf q}^2$. However, when taking the Fourier transform of $\tilde {V}(\bf q)$ (see Eq.~\eqref{eq:V2}), $|{\bf q}|$ is extended to the complex plane. The non-relativistic potential is then given by an integral of ${\rm Im}({\cal M}_t)$ over positive values of the \textit{real} variable $\lambda$, which is related to $t$ by $\lambda\equiv\sqrt{t}$. Hence the $t$ variable in Eq.~\eqref{eq:At} is positive when computing the non-relativistic potential. This implies that ${\rm Im}({\cal M}_t)$ is always negative, and thus the Casimir-Polder force between nucleons induced by a vector channel is always repulsive.

Let us finally prove Property 3.
We first remark that the long distance behaviour of the $V(r)$ potential amounts to having a steep exponential in $\int_{2m}^\infty d\lambda \lambda [\tilde V] e^{-\lambda r}$, see Eq.~\eqref{eq:V2}. When this is true we are allowed to expand $[\tilde V]$ as a power series at small values of $\lambda$, hence at the point $\lambda=2m$. In order to understand what form this power series takes, let us  consider the square amplitude $|{\cal M}(N\bar N \leftrightarrow X \bar X)|^2$, which corresponds to  pair production or annihilation of $X$.  This amplitude arises from the local operators of Eq.~\eqref{eq:V2} hence it depends only on the
 center-of-mass energy $\sqrt{s}$. We extend $s$ to the complex plane.  We can always perform a power series expansion near $s=4m^2$,\footnote{ Note that the quantity $\frac{\sqrt{s-4m^2}}{4 m}=\frac{\bf q}{m}\equiv v$  taken in the center-of-mass frame is the usual velocity of the $X$ particle. It is common to say that the squared matrix-element is ``velocity-suppressed'' when \textit{e.g.} $a=0$.  The nucleons being by assumption heavier than $X$, neither production nor annihilation of $X$ can physically happen at this threshold. However, formally, nothing forbids us to  perform the expansion.
 }

\be
|{\cal M}(N\bar N \leftrightarrow X \bar X)|^2=4m_N^2\left(a+b (s-4m^2)+c (s-4m^2)^2+\ldots\right)
\ee
where the $4m_N^2$ factor is introduced for further convenience and the $a,b,c$ are dimensionful constants.
 Using the optical theorem, we obtain that
\be
{\rm Im}({\cal M}_s)=\frac{m_N^2}{4 \pi^2}\sqrt{1-\frac{4m^2}{s}}\left(a+b (s-4m^2)+c (s-4m^2)^2+\ldots \right)\,,
\ee
and crossing then gives
\be
{\rm Im}({\cal M}_t)=\frac{m_N^2}{4 \pi^2}\sqrt{1-\frac{4m^2}{t}}\left(a+b (t-4m^2)+c (t-4m^2)^2+\ldots \right)\,.
\ee
${\rm Im}({\cal M}_t) $ is  related to  $[\tilde V]$  by Eq.~\eqref{eq:Vtilde} and  $[\tilde V]$ is related to $V(r)$   by Eq.~\eqref{eq:V2}. The potential in the long range limit turns out to be\footnote{The general case is obtained similarly using the identity
\be
\int_{2m}^\infty d\lambda \lambda (\lambda^2-4m^2)^{\frac{1}{2}+p}=\left(\frac{4m}{r}\right)^{p+1}\frac{\Gamma(3/2+p)}{\sqrt{\pi}}K_{p+1}(2m r)\,.
\ee
 }
\be
V(r)=-\frac{1}{32\pi^{5/2}}e^{-2m r}\left(
a\frac{m^{1/2}}{r^{5/2}}+b\frac{6m^{3/2}}{r^{7/2}}+c\frac{60m^{5/2}}{r^{9/2}}+\ldots
\right)\,.
\ee
We can see that an extra factor of $1/r$ in $V(r)$ is associated to each factor of $s-4m^2$ in the expansion of $|{\cal M}(N\bar N \leftrightarrow X \bar X)|^2$.


\section{Fifth force searches}\label{sec:forces}

 This section describes how to interpret the results of a number of experiments as bounds on an arbitrary fifth force.

\subsection{Neutron scattering}

Progress in measuring the scattering of cold neutrons off nuclei have been recently made and have been used to put bounds on short-distance modified gravity, \cite{Nesvizhevsky:2004qb,Leeb:1992qf,Frank:2003ms,Watson:2004vh,Greene:2006qj, Baessler:2006vm,Nesvizhevsky:2007by,Kamiya:2015eva}.
The cold neutron scattering cross-section can be measured at zero angle by ``optical'' methods,  at non-zero angles using Bragg diffraction, or over all angles by the ``transmission'' method giving then the total cross-section \cite{Koester199165}.

In the following we adapt the analyses of \cite{Nesvizhevsky:2007by}  to the Casimir-Polder forces of Eq.~\eqref{eq:forces}. At low energies the standard neutron-nuclei interaction is a contact interaction in the sense that it can be described by a four-fermion operator ${\cal O}_{4N}=\bar N  N \bar N'  N'  $.\footnote{ As described in \cite{Nesvizhevsky:2007by}, there is also a small  electromagnetic dipole interaction, which is taken into account in the analysis and which we do not discuss here. }  New physics can in general induce both contact and non-contact contributions to the neutron-nuclei interaction. A non-contact contribution vanishes at zero momentum, while a contact contribution remains non null and can be described by ${\cal O}_{4N}$. It is convenient to introduce the scattering length \be  \sqrt{\frac{\sigma({\bf q})}{4\pi}}\equiv l({\bf q})=l_{\rm std}^{\rm C}+l_{\rm NP}^{\rm C}+l_{\rm NP}^{\rm NC}({\bf q})\,, \ee where the $l_{\rm std}^{\rm C}$, $l_{\rm NP}^{\rm C}$ local terms are independent of momentum transfer ${\bf q}$ and $l_{\rm NP}^{\rm NC}({\bf q})$, which satisfies $l_{\rm NP}^{\rm NC}({\bf q}=0)= 0$,  is the non-contact contribution.
 The
$l_{\rm NP}^{\rm NC}({\bf q})$ term contains the Casimir-Polder force (see Sec.~\ref{sec:properties}), and log terms of the form $|{\bf q}|^{2n}\log(m/\Lambda)$.
The new physics contribution $l_{\rm NP}({\bf q})$ is related to the scattering potential $\tilde V$ by $l_{\rm NP}({\bf q})=2m_N \tilde V({\bf q})$, which is just the Born approximation.
For the forces described in Eq.~\eqref{eq:forces}, the new physics contributions are given by
\bea
l_a({\bf |q|^2})  &=& \frac{m_N}{16\pi^2\,\Lambda^2}f_0 \,, \\
l_b({\bf |q|^2})  &=& \frac{m_N}{4\pi^2\,\Lambda^4} \left(m^2f_0+{\bf |q|^2} f_1 \right) \,, \\
l_c({\bf |q|^2})  &=&  \frac{m_N}{16\pi^2\,\Lambda^6}\left(\left(3m^4-\frac{m^2 {\bf |q|^2}}{2} \right)f_0 +\left( \frac{  {\bf |q|^4}}{2} - 12 m^2 {\bf |q|^2} \right)f_1+10 {\bf |q|^4} f_2 \right) \,,
\eea
where the $f_n$ are the loop functions defined Eq.~\eqref{eq:fn}.
 A convenient way to look for an anomalous interaction is to search for
$l_{\rm NP}^{\rm  NC}({\bf q})$ by comparing the scattering length obtained by different methods, using for instance $l_{\rm Bragg}-l_{\rm opt}$, $l_{\rm tot}-l_{\rm opt}$. This approach eliminates the contact contributions $l_{\rm std}^{\rm C}$ and  $l_{\rm NP}^{\rm C}$, and is therefore only sensitive to $l_{\rm NP}^{\rm NC}({\bf q})$.
\begin{itemize}
\item \textit{Optical + Bragg} \\
One approach is to  compare the forward and backward scattering lengths measured respectively by optical and Bragg methods. Using the analysis from \cite{Nesvizhevsky:2007by}, one has a 95\%\,CL bound
\be
\frac{1}{2m_N}\left(l_i(0)- l_i( k^2_{\rm Bragg})\right) <  (0.01\,{\rm fm})^2\,,
\ee
with $k_{\rm Bragg} = 2$~keV.

\item \textit{Optical + Total cross-section} \\
The total cross-section measured by the transmission method provides the average scattering length
\bea
\bar l_i(k)&=& \frac{1}{2} \int_0^\pi d\theta \sin(\theta) l_i(4 k^2\sin^2(\theta/2) )\,.
\eea
Using information from optical method measurement, we have the 95\%\,CL bound
\be
  l_i(0)-\bar l_i(k_{\rm ex})< 6\cdot 10^{-4}\,{\rm fm}\,,
\ee
with $k_{\rm ex}=40$~keV.
\end{itemize}
For both methods, a  dependence on the $|{\bf q}|^{2n}\log(m/\Lambda)$ remains, which turns out to be mild in practice. Hence our results are still approximatively independent of the local four-nucleon operators - which are fixed by the unspecified UV completion (see Sec.~\ref{sec:properties}).




\subsection{Molecular spectroscopy}

Impressive progress on both the experimental
\cite{Niu201444,Biesheuvel:2016azr,UbachsAPB, Balin2011, Hori2011,PhysRevLett.98.173002,PhysRevLett.108.183003,Ubachs09}  and the theoretical \cite{Karr14,PhysRevLett.113.023004,Karr2016,KHK14,PhysRevA.76.022106,PhysRevA.82.032509, 0953-4075-37-11-010,PhysRevA.74.052506,PhysRevA.77.042506,PhysRevA.77.022509,doi:10.1021/ct900391p,Pachu11}
 sides of precision molecular spectroscopy  have been accomplished in the past decade,
opening the possibility of searching for extra forces below the \AA\,scale using  transition frequencies of well-understood simple molecular systems. Certain of these results have recently been used to bound short distance modifications of gravity, see Refs.\,\cite{PhysRevD.87.112008,Salumbides:2013dua,Ubachs17,Ubachs20161}.

The most relevant systems for which both precise measurements and predictions are available are the hydrogen molecule H$_2$, the molecular hydrogen-deuterium ion HD$^+$, the antiprotonic helium $\bar p\,^4$He$^+$ and muonic molecular deuterium ion $dd\mu^+$, where $d$ is the deuteron. These  last two systems are exotic in the sense that a heavy particle (namely $\bar p$ and $\mu^-$ respectively) has been substituted for an electron. As a result the internuclear distances are reduced, providing a sensitivity to  forces of shorter range, and thus to heavier dark particles.

 The presence of an extra force shifts the energy levels by  \be\Delta E_i = \int d^3 {\bf r}\, \Psi^*(r) V_i(r) \Psi(r)\ee  at first order in perturbation theory. We have computed these  energy shifts for the transitions between the  $(\nu=1,J=0)-(\nu=0,J=0) $ states for $H_2$, the $(\nu=4,J=3)-(\nu=0,J=2) $ states of HD$^+$, the  $(m=33,l=32)-(m=31,l=30) $ states of $\bar p \,^4$He$^+$, and the binding energy of the $(\nu=1,J=0)$ state of $dd\mu^+$ using the wave functions given in \cite{Salumbides:2013dua,Ubachs17}.
For the quantum states considered here, the typical internuclear distances are $\sim 1$\,\AA\,for H$_2$ and HD$^+$, $\sim 0.2$\,\AA\,for $\bar p\,^4$He  and $\sim 0.005-0.08$\,\AA\,for $dd\mu^+$.


The bounds on the extra forces can then be obtained by asking that $\Delta E$ be smaller than the combined (theoretical + experimental) uncertainties $\delta E$. These uncertainties are given in Tab.~\ref{tab:deltaE} (see references for details).

\begin{table}
\center
\begin{tabular}{|c|c|c|c|c|}
\hline
 & H$_2$ & HD$^+$  & $\bar p\,^4$He & $dd\mu^+$ \\
 \hline
 $\delta E$ & 
 $0.7$\,meV\,\, \cite{Salumbides:2013dua} &
 $3.9$\,neV\,\, \cite{Salumbides:2013dua} &
 $0.33$\,neV\,\, \cite{PhysRevD.87.112008} &
 $3.3$\,neV\,\, \cite{Ubachs17} \\
 \hline
\end{tabular}
\caption{Combined uncertainties for molecular spectroscopy measurements. \label{tab:deltaE}}
\end{table}


\subsection{Experiments with effective planar geometry}

A variety of experiments searching for new forces at  sub-millimeter scales are measuring the attraction between two dense objects with typically planar or spherical geometries. Whenever the distance between the objects is small with respect  to their size, these objects can be effectively approximated as infinite plates, and the force becomes proportional to the potential energy between the plates. This is the   Proximity Force (or Derjaguin's) Approximation \cite{bordag2009advances}.
An important subtlety is that most of the experiments are using objects coated with various layers of  dense materials, that should be taken into account in the computation of the force. We thus end up with calculating the potential between two plates with various layers of density for each. The effective plane-on-plane geometries are summarized in Tab.~\ref{tab:planes}. It is convenient to describe all these configurations at once using a piecewise mass density function describing $n$ layers over a bulk with density $\rho$,
\be\gamma_n(z)=
\begin{cases}
\rho_n \quad &{\rm if}\quad 0<z<\Delta_n \\
\rho_{n-1} \quad &{\rm if}\quad 0<z<\Delta_n+\Delta_{n-1}
\\ \,\,\vdots
\\ \rho \quad &{\rm if}\quad z>\sum_i^n\Delta_i \,.
\end{cases}\,.
\ee
In this notation, the  layer labelled $n$ is the closest to the other plate. The potential between an infinite plate of density structure $\gamma_a(z)$ and a plate with area $A$ and density structure $\gamma_b(z)$ at a distance $s$ is then given by
\be
V_i^{\rm plate}=2\pi A \int_0^\infty d\rho\, \rho \int_0^\infty dz_a \gamma(z_a) \int_0^\infty  dz_b \gamma(z_b) \, V_i(\sqrt{\rho^2+(s+z_a+z_b)^2})\,. \label{eq:Vplane}
\ee

\begin{table}[t]
\center
\resizebox{\columnwidth}{!}{\begin{tabular}{|c||c|c|c|c||c||c|c|c|c|c|}
\hline
Experiment& \multicolumn{4}{c||}{Plane $a$} & Separation & \multicolumn{5}{c|}{Plane $b$}   \\
 \hline
Stanford \cite{Smullin:2005iv} & \multicolumn{3}{c|}{$-$ } & Au, $30$ $\mu$m  & $25$ $\mu$m & \multicolumn{5}{c|}{Au }  \\
\hline
\multirow{2}{*}{IUPUI \cite{Chen:2014oda}}  & \multicolumn{2}{c|}{\multirow{2}{*}{Sap. }}    & \multirow{2}{*}{Cr, $10$ nm }  &  \multirow{2}{*}{Au, $250$ nm }    & \multirow{2}{*}{$[30-8000 ]$ nm } &  \multirow{2}{*}{Au, $250$ nm } & \multirow{2}{*}{Cr, $10$ nm } & Si, $2.1$ $\mu$m & \multicolumn{2}{c|}{\multirow{2}{*}{SiO$_2$ } }      \\
 & \multicolumn{2}{c|}{} & & & & & & Au, $2.1$ $\mu$m &  \multicolumn{2}{c|}{} \\
\hline
Lamoreaux \cite{Lamoreaux:1996wh,bordag2009advances} & \multicolumn{2}{c|}{ SiO$_2^{(2.23)}$ } & Cu, $0.5$ $\mu$m  & Au, $0.5$ $\mu$m  & $[0.6,6]$ $\mu$m & Au, $0.5$ $\mu$m & Cu, $0.5$ $\mu$m & \multicolumn{3}{c|}{ SiO$_2^{(2.40)}$ }  \\
\hline
AFM \cite{Fischbach:2001ry,bordag2009advances} & \multicolumn{3}{c|}{ Polystyrene }   & Au$^{(18.88)}$, $86.6$ nm  & $[62,350]$ nm &  \multicolumn{2}{c|}{Au$^{(18.88)}$, $86.6$ nm}  &  \multicolumn{3}{c|}{ Sap. }   \\
\hline
$\mu$-oscillator \cite{Decca:2007yb,Decca:2007jq,bordag2009advances} & \multicolumn{2}{c|}{ Sap.$^{(4.1)}$ } & Cr, $10$ nm  & Au, $180$ nm  & $[180,450]$ nm &  Au, $210$ nm &  Cr, $10$ nm &  \multicolumn{3}{c|}{ Si }   \\
\hline
\multirow{2}{*}{Casimirless \cite{Decca:2005qz,Decca:2007jq,bordag2009advances}}  & \multicolumn{2}{c|}{\multirow{2}{*}{ Sap. }}    & \multirow{2}{*}{ Cr, $1$ nm }  &  \multirow{2}{*}{ Au, $200$ nm }    & \multirow{2}{*}{$[150,500 ]$ nm } &  \multirow{2}{*}{Au, $150$ nm } & \multirow{2}{*}{Pt, $1$ nm }  & Ge, $200$ nm    & \multirow{2}{*}{Ti, $1$ nm }  &   \multirow{2}{*}{Si }  \\
 & \multicolumn{2}{c|}{} & & & & & & Au, $200$ nm  &  & \\
\hline
\end{tabular}}
\caption{
Summary of the fifth forces experiments with effective planar geometry used in this work. The reported densities which differ from the nominal ones given in Tab.~\ref{tab:densities} are indicated in parenthesis.
}\label{tab:planes}
\end{table}

\begin{table}[t]
\center
\begin{tabular}{|c|c|c|c|c|c|c|c|c|c|}
\hline
& Pol.  & SiO$_2$ & Si & Sap. & Ti  & Ge & Cr & Cu & Au \\
\hline
$\rho\,\,$ [ g cm$^{-3}$] & 1.06 & 2.23 & 2.33   & 3.98 & 4.51  & 5.32 & 7.14 & 8.96 & 19.32 \\
\hline
$\rho\,\,$ [$10^6 \cdot$ keV$^4$] & 4.75 & 9.99 & 1.04 & 1.78 &2.02 &2.38 &3.20 &4.01 &8.66 \\
\hline
\end{tabular}
\caption{Densities of the materials used in the fifth force experiments listed in Tab.~\ref{tab:planes}}  \label{tab:densities}
\end{table}

In practice, most of these sub-millimeter experiments have released their results as bounds on a Yukawa-like force. In order to obtain consistent bounds on the strength of the Casimir-Polder forces $\Lambda$ as a function of the scalar mass $m$, we have to compare the plane-on-plane potentials from the Casimir-Polder forces to the plane-on-plane potential from the Yukawa force. Bounds on the $(\alpha, m)$ parameters of the Yukawa force can be then translated into bounds on the $(\Lambda, m)$ parameters of the Casimir-Polder forces, using the limit-setting procedure provided by each experiment.

The plane-on-plane potential for the Yukawa force is straightforward to compute analytically and reads
\be V_{\rm Yuk}^{\rm plate}=2\pi A \frac{1}{m^3} e^{- ms } K_n^a K_{n'}^b\,, \quad K_n= \rho_n + \sum_{l=1}^n (\rho_{l-1}-\rho_l)\exp\left(-m{\sum_{i=1}^l \Delta_{n-i+1} }\right)
\ee with $\rho_0=\rho$. In  the case of the Casimir-Polder forces shown in Eq.~\eqref{eq:forces}, the triple integral of Eq.~\eqref{eq:Vplane}   are much less trivial to carry on analytically. A numerical integration is however easily done.

It is worth noticing that the $z$-integrals on the Casimir-Polder potentials can be realised using a different representation for the potentials, which naturally occurs  when calculating the diagram of Fig.~\ref{fig:loop} in a mixed position-momentum space formalism - which we will extensively use in future work \cite{WIP}.

\subsection{Bouncing Neutrons}

New forces can also be probed using bouncing ultracold neutrons (i.e. neutrons with velocities of a few m/s)
\cite{Nesvizhevsky2004,Brax2011,Brax2013,Jenke2014,Pignol2015}.
The vertical motion of a neutron bouncing above a mirror nicely realizes the situation of a quantum point particle confined in a potential well,
the gravitational potential $m_N g z$ pulling the neutron down, and the mirror pushing the neutron up.
The properties of the discrete stationary quantum states for the bouncing neutron can be calculated exactly.
The wavefunction of the $k^{\rm th}$ state reads:
\begin{equation}
\psi_k(z) = C_k {\rm Ai}(z/z_0 - \epsilon_k),
\end{equation}
where ${\rm Ai}$ is the Airy function, $\epsilon_k$ is the sequence of the negative zeros of ${\rm Ai}$ and $z_0 = (2m_N^2g/\hbar^2)^{-1/3} \approx 6 \ \mu$m.
The theoretical energies of the quantum states are
\begin{equation}
E_k = m_N gz_0 \epsilon_k = \{ 1.41, 2.46, 3.21, 4.08, \cdots \} \ {\rm peV}.
\end{equation}
Recently, a measurement of the energy difference $E_3-E_1$ was performed at the Institut Laue Langevin in Grenoble using a resonance technique \cite{Cronenberg2015}.
The result is in agreement with the theoretical predictions.
From this experiment a bound can be set on any new force which would modify the energy levels, the experimental precision being
\begin{equation}
\label{ExpQuantumBouncer}
\delta (E_3-E_1) < 10^{-14} \ {\rm eV}.
\end{equation}

Let us calculate the energy shift due to the new Casimir-Polder Dark force.
The additional potential of a neutron at a height $z$ above a semi-infinite glass mirror is given by
\begin{equation}
V_{i,z}(z) = 2\pi \frac{\rho_{\rm glass}}{m_N} \int_z^\infty dz' \int_0^\infty \rho d \rho V_{i}(r)
\end{equation}
where $\frac{\rho_{\rm glass}}{m_N} = 10^{10} \ {\rm eV}^3$ is the number density of nucleons in the glass,
$V_{i}(r)$ is the potential between the neutron and one nucleon at a distance $r = \sqrt{\rho^2 + z'^2}$.
The double integral in the expression of the potential can be simplified to a single integral:
\begin{equation}
V_{i,z}(z) = 2\pi \frac{\rho_{\rm glass}}{m_N} \int_z^\infty r(r-z) V_{i}(r) dr.
\end{equation}
In the case of the potentials $V_a$ and $V_b$, the integrals cannot be calculated analytically.
However we found suitable analytical approximations having the correct asymptotic behaviour at zero and infinite height:
\begin{eqnarray}
\nonumber
V_{a,z}(z) & = & - \frac{\rho_{\rm glass}}{m_N} \frac{1}{32 \pi^2 \Lambda^2} \int_{2mz}^\infty \frac{u-2mz}{u} K_1(u) du  \\
& \approx & - \frac{\rho_{\rm glass}}{m_N} \frac{1}{32 \pi^2 \Lambda^2} \frac{K_0(2mz)}{1+2mz},
\end{eqnarray}
and
\begin{eqnarray}
\nonumber
V_{b,z}(z) & = & \frac{\rho_{\rm glass}}{m_N} \frac{m^2}{4 \pi^2 \Lambda^4} \int_{2mz}^\infty \frac{u-2mz}{u^2} K_2(u) du \\
& \approx & \frac{\rho_{\rm glass}}{m_N} \frac{m^2}{4 \pi^2 \Lambda^4}  \frac{K_1(2mz)}{2mz (3+2mz)}.
\end{eqnarray}
The approximate expressions have a relative precision of better than 50 \% for $V_{a,z}$ and better than 3 \% in the case of $V_{b,z}$, for all values of $z$. The case of $V_{c,z}$ remains to be done.

Using the approximate expressions, we have computed the shift in the energy levels of the neutron quantum bouncer using first order perturbation theory:
\begin{equation}
\delta E_k = \int_0^\infty |\psi_k(z)|^2 V_z(z) dz.
\end{equation}
The bounds on the extra forces $V_a$ and $V_b$ as a function of the mediator mass $m$ are obtained from the experimental constraint (\ref{ExpQuantumBouncer}).
They are reported in Figs.~\ref{fig:1},\,\ref{fig:2}.

\subsection{Moon perihelion precession}

\label{sec:moon}

The existence of a fifth force at astrophysical scales would imply a slight modification of planetary motions. Any such fifth force can be treated perturbatively whenever it is small with respect to gravity at the distance between the two bodies. The modification of the equation of motion implies, among other effects,  an anomalous precession of the perihelion of the orbit. In the case of the Moon, this precession is experimentally measured to high precision  by lunar laser ranging experiments \cite{PhysRevLett.93.261101}.

 The fundamental Casimir-Polder forces of Eq.\eqref{eq:forces} are between two nucleons. For macroscopic bodies, the potentials are given by $\frac{m_1m_2}{m_N^2}V_i$.  Let us calculate the planetary motion in the presence of these new forces. We follow  the formalism of Ref.~\cite{Fischbach:1999bc}.  The radial component of the Casimir-Polder forces between Earth and Moon is given by  $F_i(r)=-\frac{m_{\tiny{\leftmoon}} m_{ \Earth}}{m_N^2}\partial_r V_i(r)$.  Introducing $u=\frac{1}{r}$, the Earth-Moon orbital equation reads
\be
\frac{d^2 u}{d\theta^2}+u = \frac{m_{\tiny{\leftmoon}}^2}{L^2 u^2}\left( m_{\tiny{\leftmoon}} m_{ \Earth} G u^2 - F_i(1/u)\right)\,,
\ee
where $L\equiv m_{\tiny{\leftmoon}}r^2 \frac{d\theta}{dt}$ is the conserved angular momentum and  the first term in the parenthesis is the gravitational force. The solution of the unperturbed equation reads \be u(\theta)=u_{\tiny{\leftmoon}}\left(1+\epsilon \cos(\theta-\theta_0)\right)\,,\quad u_{\tiny{\leftmoon}}=
\frac{m_{\tiny{\leftmoon}}^3 m_{\Earth} G}{L^2 }  \ee where $\epsilon$ is the excentricity ($\epsilon=0.0549$ for the Moon), $\theta_0$ indicates the perihelion of the ellipse, and the major semiaxis $a_{\tiny{\leftmoon}}$ is given by $a_{\tiny{\leftmoon}}^{-1}=u_{\tiny{\leftmoon}}(1-\epsilon^2)$. At first order in perturbation theory, the extra force is just as a constant, $F_i(1/a_{\tiny{\leftmoon}})$, which only modifies $u_{\tiny{\leftmoon}}$, the overall size of the orbit. At second order in perturbation theory, one has
\be
F_i(1/u)=F_i(1/a_{\tiny{\leftmoon}})+\left(u-\frac{1}{a_{\tiny{\leftmoon}}}\right) \frac{\partial F_i(1/u)}{\partial u}\bigg|_{u=1/a_{\tiny{\leftmoon}}}\,.
\ee
The  term linear in $u$ modifies the frequency of the orbit on the left-hand side of the equation of motion. The motion  is now given by
\be
u(\theta)=u_{\tiny{\leftmoon}}\left(1+\epsilon \cos \omega(\theta-\theta_0)\right)(1+\ldots)\,, \quad \omega^2 = 1+ \frac{u_{\tiny{\leftmoon}} a_{\tiny{\leftmoon}}^4}{G m_N^2} \partial^2_r V_i(r)|_{r=a_{\tiny{\leftmoon}}}
\ee
where the ellipsis denotes irrelevant corrections to the overall magnitude of the orbit. Having $\omega \neq 1$ implies a precession of the perihelion, which can be seen using $\cos\omega(\theta-\theta_0)=\cos\omega\left(\theta-\theta_0+\frac{2\pi n }{\omega}\right)$. The precession angle between two rotations is finally given by \be \delta \theta_i = -\pi \frac{a_{\tiny{\leftmoon}}^3}{G m_N^2(1-\epsilon^2)}  \partial^2_r V_i(r)|_{r=a_{\tiny{\leftmoon}}}\,.  \ee
We apply this general formula to the Casimir-Polder potentials of Eq.~\eqref{eq:forces}.
Interestingly, the $V_a$ and $V_c$ potentials, which are attractive, induce an advance of the perihelion while $V_b$, which is repulsive, induces a delay of the perihelion.

The Moon precession angle is constrained by lunar laser ranging experiments. Other well-understood perturbations induce Moon's orbit precession: the quadrupole field of the Earth, other bodies of the solar system, and general relativity. Once all these effects are taken into account, one obtains a bound on an extra,  anomalous precession angle. Following Ref.~\cite{Adelberger:2003zx},  an experimental limit from lunar laser ranging  is given as \be\delta \theta_i< 2\pi \times 1.6 \cdot 10^{-11}\,. \ee

%
%
%
%
%
%
%
%
%

\section{Bounds on  forces from dark  scalars}\label{sec:results}

\begin{figure}
\centering
\includegraphics[width=0.77\textwidth]{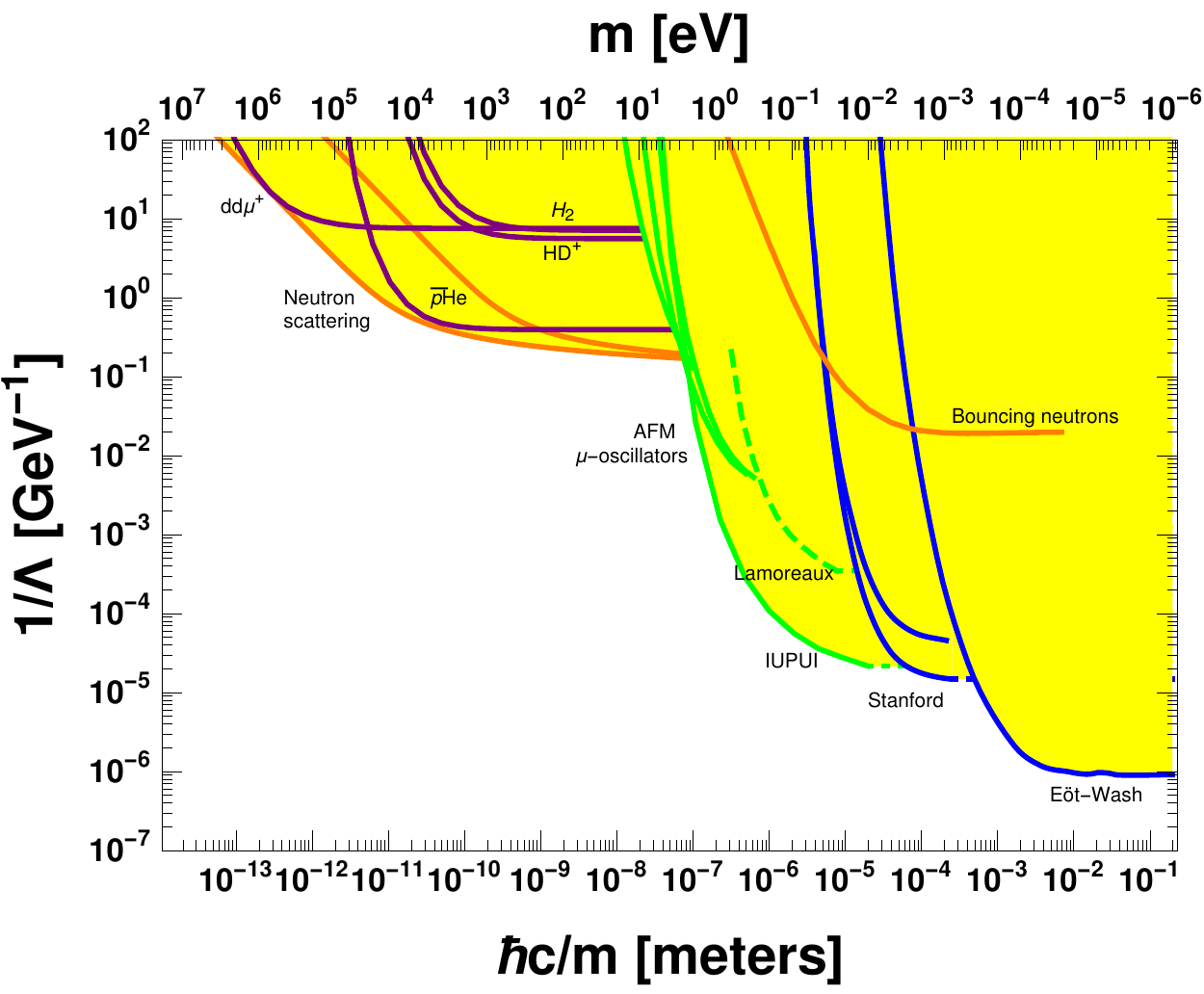}
\caption{\label{fig:1}
Bounds on a scalar coupled to nucleons via the ${\cal O}_a$ interaction. The yellow region is excluded at $95\%$~CL.  See Sec.~\ref{sec:forces} for details on exclusion regions.
}
\end{figure}

Let us apply the experimental bounds obtained in Sec.~\ref{sec:forces} to the Casimir-Polder forces from a dark scalar given in Eq.~\eqref{eq:forces}.

It is instructive to understand first qualitatively the landscape of exclusion regions on the Casimir-Polder forces. Let us consider  the exclusion regions for the Yukawa force (see \textit{e.g.} \cite{Adelberger:2009zz}). Starting from large scales, the reach of the experiment starts to decrease very steeply below the scale of the E\"ot-Wash experiment,  at roughly $\lambda<O(10^{-4}$~m)  down to atomic scales.  In this region of $\lambda$, the bound on the strength of the Yukawa force  $\alpha$ scales very roughly as $\alpha<  10^{-22} \, \left(\frac{1\,  {\rm m}}{\lambda}\right)^5  $, demonstrating the increasing difficulties in measuring forces at small distances.
The Casimir-Polder forces    behave as $1/r^n$ with $n \geq 3$ at short distance. This has the crucial implication that  the constraints from short distance experiments will gain importance and those from long distance will lose importance compared to the exclusion regions on the Yukawa-like force.  In particular, one can expect the E\"ot-Wash bound to dominate over the bounds from all experiments at larger   scale, to the  possible exception of lunar laser ranging. Moreover, when $n=7$, the decrease of sensitivity in
\begin{figure}[p!]
\centering
\includegraphics[width=0.77\textwidth]{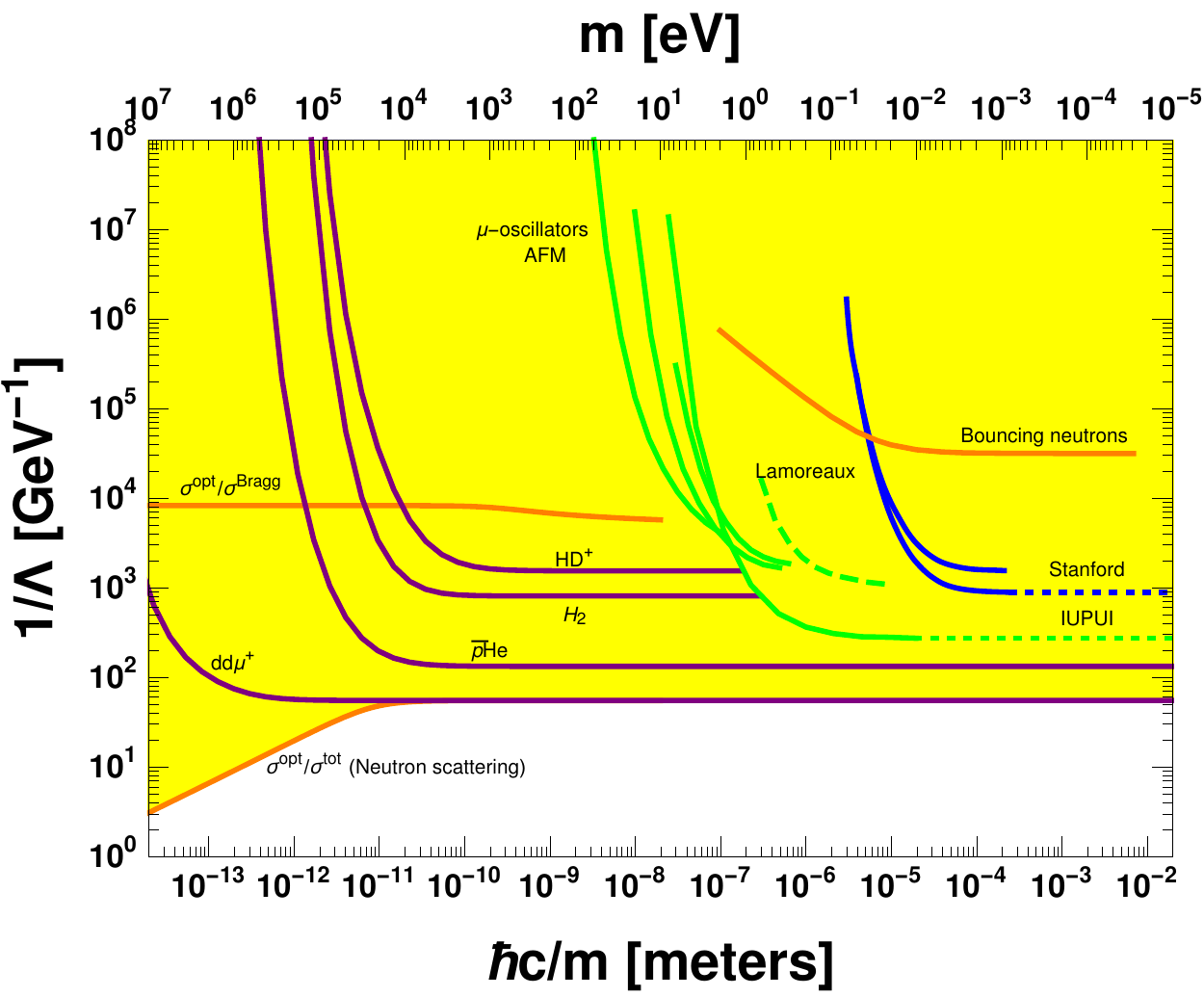}
\caption{\label{fig:2}
Bounds on  a scalar coupled to nucleons via the ${\cal O}_b$ interaction. Same conventions as Fig.~\ref{fig:1}. }
\end{figure}
\begin{figure}[p!]
\centering
\includegraphics[width=0.77\textwidth]{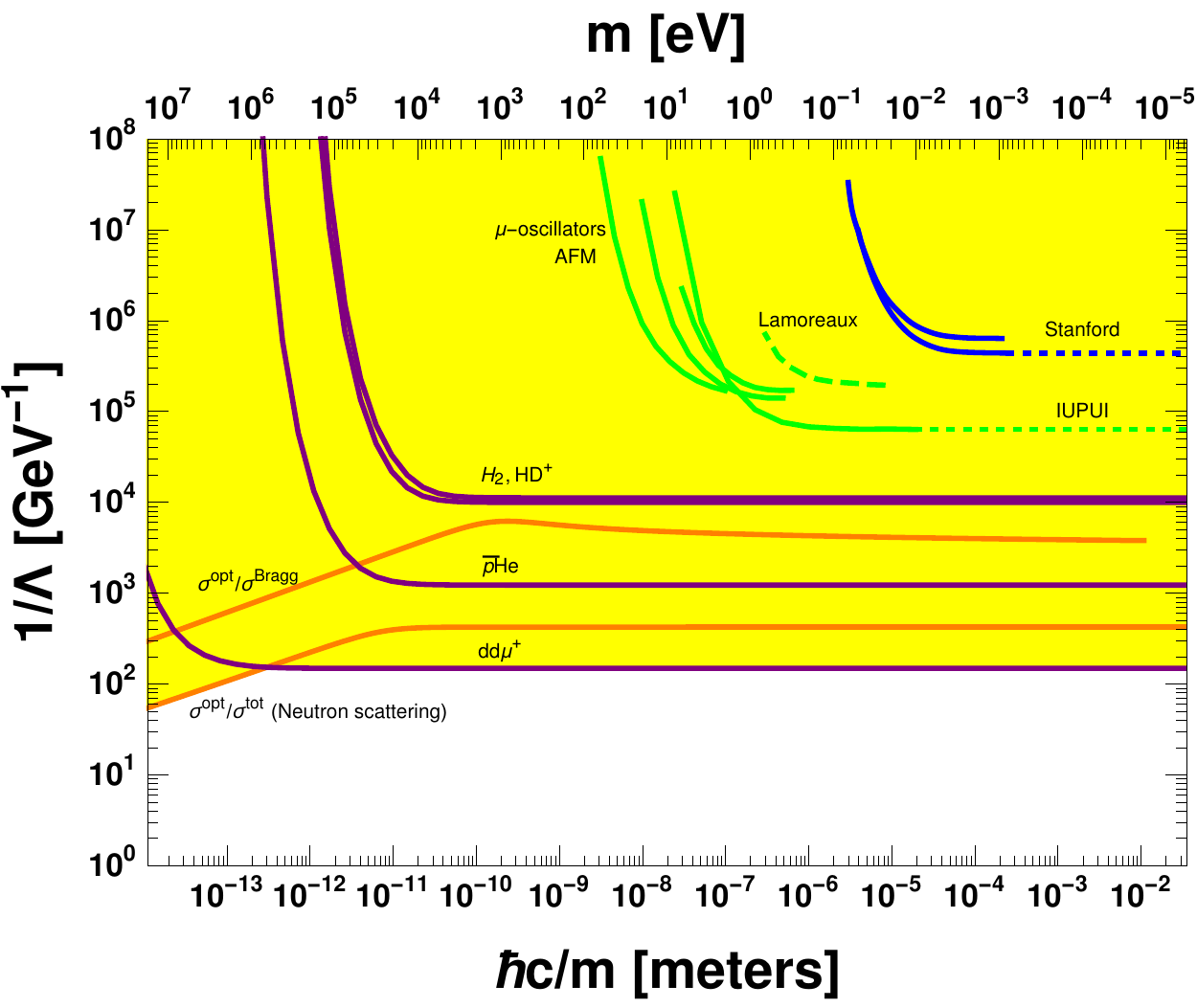}
\caption{\label{fig:3}
Bounds  on a scalar coupled to nucleons via the ${\cal O}_c$ interaction. Same conventions as Fig.~\ref{fig:1}.}
\end{figure}
\clearpage
\noindent $\lambda^{-5}$ is expected to be overwhelmed by the increase of the force in $r^{-7}$, implying that bounds from the   experiments at the smallest scales (from neutron scattering and molecular spectroscopy) dominate over all the bounds from higher distances.

The exclusion regions for the $V_a$, $V_b$, $V_c$ Casimir-Polder potentials  are respectively  presented in Figs.~\ref{fig:1},\,\ref{fig:2},\,\ref{fig:3}.
For the $V_a$ potential, we obtain that the E\"ot-Wash bound is the dominant one for $\lambda> 10^{-3}$\,m.
For both $V_b$ and $V_c$ potentials, we obtain indeed an inversion in the hierarchy of bounds. The two leading bounds turn out to be from the  $dd \mu^+$ molecular ion and from the neutron scattering bound combining optical and total cross-sections.  This fact can be taken as an incentive to pursue and develop such small scale experiments.

Interestingly, for $V_a$, the bound from antiprotonic helium $\bar p$\,He$^+$  is stronger than the bound from the $dd \mu^+$  ion, while it is not the case for the $V_b$, $V_c$ potentials. This feature comes from the fact that the wave function of the $dd \mu^+$ ground state has a large tail towards short distances. This tail enhances the contribution of the potentials which grow faster at small distance, hence the  $dd \mu^+$ bound gets favored with respect to the $\bar p$\,He$^+$ bound for $V_b$ and even more $V_c$. The leading bound being either $dd \mu^+$ or $\bar p$\,He$^+$ depending on the potential, further studies (both theoretical and experimental) in both systems should definitely be encouraged.

Using the calculation given in \ref{sec:moon}, we obtain that limits from lunar laser ranging are indeed subleading.  At zero mass,  the bounds on $\Lambda$ for the $V_a$, $V_b$, $V_c$ potentials are found to be respectively $\Lambda> 2$\,GeV, $6\cdot 10^{-5}$~eV, $2\cdot 10^{-8}$~eV. All these bounds are overwhelmed by stronger ones from shorter distance experiments.



\section{Conclusions}

There are many motivations---including Dark Matter and Dark Energy---for speculating on the existence of a dark sector containing particles with a bilinear coupling to  the Standard Model particles.   Whenever one of the dark particles is light enough and couples to nucleons in a spin-independent way, it induces forces of the Casimir-Polder type, that are potentially accessible by fifth force experiments across many scales.  The short and long-range behaviours of these forces as well as their sign can all be understood and predicted using dimensional analysis and the optical theorem.
We provide a comprehensive (re)interpretation of bounds from neutron scattering to the Moon perihelion precession, applicable to any kind of potential. We then focus on the case of a scalar with a variety of couplings to nucleons, generating forces with $1/r^3$, $1/r^5$, $1/r^7$ short-distance behaviours.
It turns out that forces in $1/r^5$, $1/r^7$ are best constrained by neutron scattering and molecular spectroscopy, which provides extra motivation to pursue these kind of low-scale experiments. Implications for Dark Matter searches have been discussed in Ref.~\cite{Fichet:2017bng}.

\section*{Acknowledgements}

This work is supported in part by the EU Horizon 2020 research and innovation programme under the Marie-Sklodowska grant No. 690575. This article is based upon work related to the COST Action CA15117 (CANTATA) supported by COST (European Cooperation in Science and Technology).  SF work was supported by the S\~ao Paulo Research Foundation (FAPESP) under grants \#2011/11973 and \#2014/21477-2.
\\
\\
\\

\appendix

\section{Calculation of the potentials}

This appendix contains details of the computation for the potentials in Eq.~\eqref{eq:forces} and those given in  Ref.~\cite{Fichet:2017bng}. The full set of operators considered is
\bea \nonumber \label{eq:Leff}
{\cal O}^0_a&=&\frac{1}{\,\Lambda} \bar N N |\phi|^2  \,,\quad
{\cal O}^0_b=  \frac{1}{\,\Lambda^2}\, \bar N \gamma^\mu N \phi^* i\overleftrightarrow\partial_\mu  \phi
  \,,\quad
{\cal O}^0_c=\frac{1}{\,\Lambda^3} \bar N N \partial^\mu \phi^* \partial_\mu\phi \,,
\\
  {\cal O}^{\nicefrac{1}{2}}_a&=&\frac{1}{\,\Lambda^2} \bar N N \bar \chi \chi \,, \quad  {\cal O}^{\nicefrac{1}{2}}_b=\frac{1}{\,\Lambda^2} \bar N \gamma^\mu N \bar \chi \gamma^\mu \chi \,,\quad
  {\cal O}^{\nicefrac{1}{2}}_c=\frac{1}{\,\Lambda^2} \bar N \gamma^\mu  N \bar \chi \gamma^\mu \gamma^5\chi \,,
\\ \nonumber
{\cal O}^{1}_a&=&\frac{1}{\Lambda^3} \bar N N | m X^\mu+\partial^\mu\pi|^2  \, , \quad
{\cal O}^{1}_b=\frac{1}{\Lambda^2} \,2 \bar N \gamma^\mu N \,{\rm Im}(X_{\mu\nu }X^{\nu *}
+ \partial^\nu(X_\nu X^*_\mu) +\partial^\mu \bar c c^*
)\,,  \\ \nonumber
{\cal O}^{1}_c&=&\frac{1}{\Lambda^3} \bar N N |X^{\mu\nu}|^2 \,, \quad {\cal O}^{1}_d=\frac{1}{\Lambda^3} \bar N N X^{\mu\nu}\tilde X^*_{\mu\nu}\,.
\eea
A dark particle of spin $0$, $1/2$, $1$ is denoted by $\phi, \chi, X$. $\pi$ and $c, \bar c$ are respectively the Goldstone bosons and ghosts accompanying $X$.
At that point the dark particle can be self-conjugate (real scalar or vector, Majorana fermion) or not (complex scalar or vector, Dirac fermion). When $X$ is complex, so are $\pi$, $c$ and $\bar c$.
We will give the results for all cases. We introduce
\be
\eta=
\begin{cases}
0 \quad \textrm{if self-conjugate}\\
1 \quad \textrm{otherwise} \,.
\end{cases}
\ee
We calculate the loop diagram of Fig.~\ref{fig:loop} induced by each of these operators using dimensional regularisation. The matching of the effective theory with the UV theory being done at the scale $\Lambda$, we can readily identify the divergent integrals as  (see \cite{peskin1995introduction,Alam:1997nk})~\footnote{The running of the Wilson coefficients is taken into account at leading-log order with this method. }
\be
\int \frac{d^4 l}{(2\pi)^4}\frac{1}{(l^2-\Delta)^2} \rightarrow \frac{-i}{(4\pi)^2}\log(\Delta/ \Lambda^2)\,,
\label{eq:L0}
\ee
\be
\int \frac{d^4 l}{(2\pi)^4}\frac{l^2}{(l^2-\Delta)^2} \rightarrow \frac{-2\,i }{(4\pi)^2}\Delta\log( \Delta/ \Lambda^2)\,,
\label{eq:L1}
\ee
\be
\int \frac{d^4 l}{(2\pi)^4}\frac{(l^2)^2}{(l^2-\Delta)^2} \rightarrow \frac{-3\,i }{(4\pi)^2}\Delta^2\log( \Delta/ \Lambda^2)\,.
\label{eq:L2}
\ee
From these amplitudes, the discontinuities in the non-relativistic scattering potential $\tilde V$ are given by Eq.~\eqref{eq:NRV} and are found to be
\bea
\left[ \tilde V^0_a\right]&=& 2^\eta \frac{\left[f_0\right]}{32 \pi^2\, \Lambda^2}  \\ \nonumber
\left[ \tilde  V^0_b\right]&=&\eta \frac{m^2 \left[f_0\right]-\lambda^2 \left[f_1\right]}{8 \pi^2\, \Lambda^4} \\ \nonumber
\left[\tilde  V^0_c\right]&=&2^\eta  \frac{(6m^4 +m^2\lambda^2) \left[f_0\right] -
(24m^2 \lambda^2 +\lambda^4) \left[f_1\right]+
20\lambda^4\, \left[f_2\right]
}{64   \pi^2\, \Lambda^6}\\ \nonumber
\left[ \tilde  V^{1/2}_a\right]&=&2^\eta \frac{3(\lambda^2\left[f_1\right]-m^2 \left[f_0\right] )}{8 \pi^2\, \Lambda^4} \\ \nonumber
\left[ \tilde  V^{1/2}_b\right]&=&\eta \frac{ -\lambda^2 \left[f_1\right]}{2 \pi^2\, \Lambda^4} \\ \nonumber
\left[ \tilde  V^{1/2}_c\right]&=&2^\eta \frac{m^2 \left[f_0\right] -\lambda^2 \left[f_1\right]}{4 \pi^2\, \Lambda^4} \\ \nonumber
\left[\tilde  V^1_a\right]&=&2^\eta  \frac{(6m^4 -m^2\lambda^2) \left[f_0\right] -
(12m^2 \lambda^2 +\lambda^4) \left[f_1\right]+
20\lambda^4\, \left[f_2\right]
}{64   \pi^2\, \Lambda^6}
\\ \nonumber
\left[\tilde  V^1_b\right]&=&\eta \frac{ (8m^2+5\lambda^2) \left[f_0\right]-10\lambda^2\left[f_1\right]}{16 \pi^2\, \Lambda^4}
\\ \nonumber
\left[\tilde  V^1_c\right]&=&2^\eta  \frac{(9m^4 +3m^2\lambda^2) \left[f_0\right] -
(36m^2 \lambda^2 +3\lambda^4) \left[f_1\right]+
30\lambda^4\, \left[f_2\right]
}{8   \pi^2\, \Lambda^6}\\ \nonumber
\left[\tilde  V^1_d\right]&=&2^\eta  \frac{3(\lambda^4 \left[f_1\right] - \lambda^2 m^2 \left[f_0\right])
}{8   \pi^2\, \Lambda^6}\\ \nonumber
\eea
where the discontinuities of $f_{0,1,2}$ are given in Eq.~\eqref{eq:fn}.
These loop functions are
 explicitly given by
\bea
f_0(m^2, q^2, \Lambda)= 2 L\left(\frac{4m^2}{q^2}\right)+\log\left(\frac{m^2}{\Lambda^2}\right)  \\
f_1(m^2, q^2, \Lambda)= \frac{2m^2 +q^2}{3 q^2} L\left(\frac{4m^2}{q^2}\right)+
\frac{1}{18}+\frac{1}{6}
\log\left(\frac{m^2}{\Lambda^2}\right) \\
f_2(m^2, q^2, \Lambda)= \frac{6m^4 +2m^2q^2 +q^4}{15 q^4} L\left(\frac{4m^2}{q^2}\right)+
\frac{13}{900}+\frac{m^2}{30 q^2}+\frac{1}{30}
\log\left(\frac{m^2}{\Lambda^2}\right)
\eea
with
\be
L(x)=\begin{cases}
\sqrt{x-1 }\arctan\left(\frac{1}{\sqrt{x-1 }}\right)-1 \quad \textrm{if} \quad x>1
\\
\sqrt{1-x }\left(i\pi+\frac{1}{2}\log\left(\frac{1+\sqrt{1-x }}{1-\sqrt{1-x }}\right)\right)-1 \quad \textrm{if} \quad x<1 \,.
\end{cases}
\ee
The $[f_n]$ discontinuities can be  obtained by noticing that $\ln \Delta= \ln (x-x_+)(x-x_-)$ where
\be
x_\pm= \frac{1}{2} \pm \frac{\sqrt{q^2-4m^2}}{2q}
\ee
has a branch cut between $x_-$ and $x_+$ and a discontinuity of $2\pi i$. This leads to
\be
[f_n]= 2\pi i \int_{x_-}^{x_+} (x(1-x))^n dx
\ee

Finally, the spatial potential is given by Eq.~\eqref{eq:V2}. The integrals over $\lambda$ needed in the last step of the calculation  are
\bea
\int_{2m}^\infty d\lambda \sqrt{\lambda^2-4m^2} e^{-\lambda r} = \frac{2m}{r} K_1(2mr) \\
\int_{2m}^\infty d\lambda \lambda^2 \sqrt{\lambda^2-4m^2} e^{-\lambda r} = \frac{8m^3}{r} K_1(2mr)+ \frac{12m^2}{r^2} K_2(2mr) \\
\int_{2m}^\infty d\lambda \lambda^4 \sqrt{\lambda^2-4m^2} e^{-\lambda r} = \frac{32m^4}{r^2} K_2(2mr)+\left(\frac{120m^3}{r^3}+\frac{32 m^5}{r}\right) K_3(2mr) \,.
\eea

\section{Amplitudes}

The one-loop amplitudes induced by the operators ${\cal O}_a$, ${\cal O}_b$, ${\cal O}_c$ are
\be
i{\cal M}_a=\frac{1}{2\Lambda^2}\bar u(p_1)u(p_2)\bar u(p_1')u(p_2') \int\frac{d^4k}{(2\pi)^4}\frac{1}{(k^2-m^2)((k+q)^2-m^2)}
 \ee
\be
i{\cal M}_b=\frac{1}{\Lambda^4}\bar u(p_1)\gamma^\mu u(p_2)\bar u(p_1') \gamma^\nu u(p_2') \int\frac{d^4k}{(2\pi)^4}\frac{(q+2k)^\mu(q+2k)^\nu}{(k^2-m^2)((k+q)^2-m^2)}
 \ee
\be
i{\cal M}_c=\frac{1}{2\Lambda^6}\bar u(p_1)u(p_2)\bar u(p_1')u(p_2') \int\frac{d^4k}{(2\pi)^4}\frac{(q.(q+k))^2}{(k^2-m^2)((k+q)^2-m^2)}
 \ee
with $q=p_1-p_2$.
These integrals can be reduced to the basis shown  in Eqs.~\eqref{eq:L0}, \eqref{eq:L1}, \eqref{eq:L2}  using textbook techniques (see \cite{peskin1995introduction}, including Feynman trick).

\bibliographystyle{JHEP}
\bibliography{biblio.bib}

\end{document}